\documentclass[12pt]{article}

\usepackage{PRIMEarxiv}

\usepackage[utf8]{inputenc} 
\usepackage[T1]{fontenc}    
\usepackage{hyperref}       
\usepackage{url}            
\usepackage{booktabs}       
\usepackage{amsfonts}       
\usepackage{nicefrac}       
\usepackage{microtype}      
\usepackage{lipsum}
\usepackage{fancyhdr}       
\usepackage{graphicx}       
\graphicspath{{media/}}     

\title{A Case for Competent AI Systems $-$ A Concept Note
}

\author{
  Kamalakar Karlapalem \\
  IIIT Hyderabad \\
  India\\
  \texttt{kamal@iiit.ac.in} \\
}

\begin{document}
\maketitle

\begin{abstract}
The efficiency of an AI system is contingent upon its ability to align with the specified requirements of a given task.
However, the inherent complexity of tasks often introduces the potential for harmful implications or adverse actions.
This note explores the critical concept of capability within AI systems, representing what the system is expected to
deliver. The articulation of capability involves specifying well-defined outcomes. Yet, the achievement of this
capability may be hindered by deficiencies in implementation and testing, reflecting a gap in the system's competency
(what it can do vs. what it does successfully).

\bigskip

A central challenge arises in elucidating the competency of an AI system to execute tasks effectively. The exploration
of system competency in AI remains in its early stages, occasionally manifesting as confidence intervals denoting the
probability of success. Trust in an AI system hinges on the explicit modeling and detailed specification of its
competency, connected intricately to the system's capability. This note explores this gap by proposing a framework for
articulating the competency of AI systems.

\bigskip

Motivated by practical scenarios such as the Glass Door problem, where an individual inadvertently encounters a glass
obstacle due to a failure in their competency, this research underscores the imperative of delving into competency
dynamics. Bridging the gap between capability and competency at a detailed level, this note contributes to advancing
the discourse on bolstering the reliability of AI systems in real-world applications.
\end{abstract}

\section {Introduction}
\textbf{The Glass Door Problem} \par Scott V. Liebman, 404 S.W.2d 288(1966) is a legal case involving James Liebman, who walked
through a glass door at the Motor Inn and injured himself. The key aspect of why James Liebman walked through the glass
door – as quoted from Scott V. Liebman, 404 S.W.2d 288(1966) is: \par
\bigskip

\emph{"We regard the evidence as being undisputed that Liebman did not see the glass door in its closed condition. There is no
suggestion that he did see it and nevertheless deliberately proceeded to walk through it anyway. One of the plaintiff's
witnesses, a business associate who was on the inside of Liebman's room, testified that he and Liebman were looking at
each other, eye to eye, as Liebman returned to the room; and Liebman proceeded as though the door were open.
\bigskip
Pictures introduced into evidence demonstrated that at night with the lights turned on inside the room, it is difficult
from the darkened outside to detect the presence of the glass door in a closed position. There was evidence that there
was a small amount of imperfection or warping in the glass; but it, too, is not easily detected. In short, this clear
glass door in a closed position at night, under the circumstances here present, would not be regarded by us, in an
abstract proposition, as an open and obvious danger of which we would charge people with knowledge."
}

\bigskip

Liebman possessed the capability to avoid injuring himself while walking through the glass door, and he believed he had
the competency to navigate the potential dangers successfully. However, despite his perceived competency, he failed to
detect the glass door, resulting in an injury. Legal cases arising from such incidents have prompted premises to adopt
preventive measures, emphasizing the need to warn individuals about glass doors and potentially enhance people's
competency to identify and avoid such hazards.

\bigskip

Metaphorically, individuals encounter various {\textquotedbl}glass doors{\textquotedbl} in their daily lives despite
their understanding (and ability to avoid mishaps) of the environment and associated risks. Human competency, often
implicit, is determined by successful outcomes of actions taken, supported by external recognition or certifications as
proof. Although individuals may perceive and deduce their surroundings accurately, actions driven by these perceptions
can sometimes lead to harm or injury.

\bigskip

Human competency is dynamic, evolving through experiences gained from performing various acts. Labels such as
{\textquotedbl}most competent surgeon, carpenter, lawyer, or President{\textquotedbl} are attributed to the competency
demonstrated in performing respective roles. Competency is a crucial criterion for task assignments, with any lapses,
especially by those deemed competent, resulting in potential legal consequences. Leveraging competency certification
enables individuals to offer services at a cost, shaping the narrative around the acts performed by individuals or
systems.

\bigskip

Applying the same principles to AI systems is imperative, acknowledging that competency remains a critical concern
despite good intentions and moral considerations. The following sections provide an overview of an AI system, exploring
micro-competency, macro-competency, the Glass Door Problem, strategies for enhancing competency, and considerations of
moral competency. Subsequent sections delve into discussions, related work, and conclusions.
 
\section{The AI System}
The AI system comprises integrated software modules and hardware components that emulate human intelligence in
performing specific tasks. Its primary objective is to enhance human capabilities and, in some instances, outperform
them. Tasks such as providing live commentary at sports events, managing traffic coordination, aiding in domestic
chores through robot assistance, achievements like Watson winning Jeopardy, and the development of self-driving cars
exemplify the diverse competencies of AI systems.

\bigskip

The AI system's definition and the extent of its capability are shaped by the environment and the system's embodiment
within it. The AI system involves (i) modeling the environment, (ii) sensing or collecting data from the environment,
(iii) decision-making based on its task at hand, and (iv) iterative execution of (i), (ii), and (iii) until the goal,
defined by the successful completion of the task, is achieved. The competency of the AI system is delineated through
these four key aspects.

\bigskip

The AI system (in a simplistic manner) can be structured into modules or hardware components, including a world module
(i), sensing module (i, ii), machine learning module (iii, iv), planning module (iii, iv), action module (iii, iv), and
interfacing module (i, ii, iii, iv). It is essential to note that an AI system can function as a part of each module,
resulting in an intricate system of integrated AI systems (subsystems). Some are explicit, while others remain
implicit, concealed within a software module or a hardware component. For instance, the control module governing hand
manipulator movement may contain an implicit, inaccessible machine learning module. In contrast, the machine learning
module for determining the optimal robot path might be explicit and visible. Such a framework aids in comprehending the
competency of the AI system.

\bigskip

Each module represents the environment and a set of algorithms embodying the AI system's functionality as tools. These
tools, operating cohesively, can be considered autonomous agents (the AI system) or specialized tools for specific
tasks. Consequently, each module or component contributes to the competence of tools and AI subsystems, ultimately
shaping the overall competency of the entire AI system.

\section[Micro{}-Competence]{Micro-Competence}
Micro-competence pertains to the proficiency of discrete components within the AI system, focusing on tools or
subsystems executing minute, atomic actions. These tools, operating as standalone modules, exhibit a certain level of
competency. Examples include Natural Language Processing (NLP) stemmers, image segmentation algorithms, manipulator arm
motors, scenario analysis modules, communication protocols for agents, decision tree algorithms, and data-collecting
sensors. Each module or component encompasses a representation, data, algorithms, and a specific competency level.

\subsection{Representation Competency}

Representation competency is pivotal as the AI system, or its tools leverage the world model to assess the system's
state and plan subsequent actions. The representation of the world model, be it through facts and predicates or
geographical coordinates like longitude and latitude, significantly influences AI systems processing. While different
tools and components may employ diverse representations, the chosen representations dictate the competency of the tools
and components.

\subsection{Data Competency}

Sensed data fundamentally drives the AI system, and the competency of the tools and components revolves around the
data's source, quality, relevance, and fitness. Issues concerning data competency encompass (a) the integrity of the
data source, (b) ensuring data quality, (c) relevance to the task at hand, and (d) fitness for use by the tools and
components. Secure and trustworthy data sources, proper data flow, and consideration of contextual relevance are
imperative to maintain data competency.

\bigskip

Data quality is contingent on the mode of data collection, whether from human or automatic sources. It involves ensuring
accuracy, mitigating noise, addressing transfer errors, and approximating data when needed. The impact of data quality
lapses on the AI system's competency is unpredictable and necessitates careful measurement.

\bigskip

Relevance of the data is critical in determining the competency of the AI system, requiring the data to align with the
tool's functionality and generate the desired output. Techniques like feature extraction, although designed to identify
relevant data, may have limitations, such as the suitability of Principal Component Analysis (PCA) for specific machine
learning tasks.

\bigskip

The fitness of the data can occur at multiple levels (1) the core algorithm expects some data with some features, and it
is unavailable, (2) the data needed for the core algorithm is not available. In this case, the system will not be able
to function. For example, the tool expects the world model as facts and predicates, but the world model is available
only in OpenStreetMap format. Fitness can be more critical in natural language processing when there is a mismatch in
vocabulary, dictionary, or even the expected parsing level. For example, for an Amazon Echo to entertain a
four-year-old toddler with some puzzles, there must be a language match between Echo and toddler and vice-versa.

\subsection{Algorithmic Competency}

Algorithmic competency is determined by the algorithms steering the AI system. These algorithms process data, and
generate outputs driving decision-making, action-taking, or planning. The algorithm's competence can range from optimal
solutions to confidence-driven outcomes influenced by evaluation parameters generated within the algorithm.
Considerations include processing time, fine-tuning with explicit parameter values, and balancing optimality and
fairness. Each algorithm contributes to the overall competency of the AI system.

\subsection{Functional Competency}

Functional competency-primarily applies to hardware components, delineating their capabilities. For instance, a wheeled
robot's functional competency might encompass specifications for maximum torque and efficient energy utilization.
Explicit utilization of functional competency is crucial for macro-level decision-making and actions by the AI system.
Agents within multi-agent or larger AI systems also possess explicit functional competency. Micro-competency
underscores the competence of individual components and modules, necessitating articulation through formal methods
conducive to reasoned understanding.

\section[Macro{}-Competency]{Macro-Competency}
Macro-competency denotes the comprehensive proficiency of the AI system at a broader scale, reflecting its loosely
integrated yet occasionally tightly bound structure influenced by data and control flows across modules and components.
The multi-agent system paradigm provides a fitting representation where explicit knowledge of each agent's behaviors
may not always be available. The overall competency at the macro level encompasses flow, action, and solution
competencies.

\subsection{Flow Competency}

Flow competency ensures the smooth operation of the AI system, requiring seamless data and control event flows. In many
production AI systems, these flows are often either hard-coded or not explicitly disclosed, complicating the
identification of system failures. The lack of explicit data and algorithmic competency disclosure for tools and
components further hampers understanding. Articulating flow competency at the macro level is crucial, and it can be
accomplished through workflows and formal models such as Petri-Nets or temporal logic.

\bigskip

For optimal functionality, data must be available to the right tool at the right time to meet processing deadlines and
facilitate decision-making. Articulating flow competency includes specifying time-related constraints necessary for the
AI system to perform effectively.

\subsection{Action Competency}

Action competency involves the independent generation and execution of actionable outputs by the AI system. It is
pivotal for systems with numerous actionable components. Understanding the range of possible actions and evaluating the
viability of actions taken is critical. Dependencies among modules and interdependent actions must be explicit to
determine the combined competency of the system at the macro level.

\bigskip

For instance, in a humanoid football team, not only must actions be decided in real-time, but there must also be
sufficient time for actions to be completed. Explicitly outlining dependencies and interdependencies is essential for
evaluating the combined competency of the system.

\subsection{Solution Competency:}

The AI system acts as a solution generator and implementer, fulfilling specific goals despite being composed of modular
components. Solution competency is influenced by micro-level competency aspects. It is limited by the least competent
module or component, often referred to as the weakest competent link, participating in the solution.

\bigskip

Explicitly articulating solution competency enables architects to enhance the overall competency by addressing the
weakest link. This articulation considers the AI system's achievable goals and how different modules and components
collaborate to achieve solutions. If the AI system comprises other AI systems, solution competency is the cumulative
competency, incorporating solution dependencies among the constituent AI systems.

\bigskip

In summary, articulating macro-competency is essential to comprehensively understand the AI system's effectiveness,
considering its intricate interplay of flows, actions, and solution-generation capabilities.

\section{The Glass Door Problem}
The Glass Door Problem introduced metaphorically through an incident where an individual attempt to walk through a glass
door, underscores a systemic challenge within the AI system. This predicament arises when a person possessing the
necessary capability encounters difficulties comprehending the environment or processing sensed data, leading to
inadvertent collisions with the glass door (showing a lack of competency). Similarly, an AI system, exemplified by
scenarios like accidents involving a self-driving car, may struggle to competently process solutions, potentially
harming others, such as a robot failing to achieve its goal or causing unintended harm.

\bigskip

Precisely articulating the system's competency is pivotal in defining the Glass Door Problem. Subsequently, as the
solution is executed, instances of failures or accidents prompt the system to identify deficiencies in its competency
within modules or components across numerous solution executions. Through ongoing audits, the competency of these
modules and components is assessed, unveiling shortcomings in the overall AI system. The audit becomes a crucial step
in fortifying the system's competency.

\bigskip

Enhancing the system's competency necessitates systematic competency analytics, aligning users' understanding of its
capabilities with its actual competency. This process mirrors how individuals identify a competent mechanic or doctor
based on personal experiences or the experiences of others. The comparison between perceived competency and actual
system competency is a foundational aspect for ongoing improvement and development of the AI system.

\section{Enhancing Competency}
The ongoing improvement of the AI system necessitates a feedback loop dedicated to monitoring and enhancing its
competency. This involves a comprehensive analysis of the system's macro-level and micro-level competencies. Modules
and components susceptible to failures, errors, and accidents undergo scrutiny at the micro-competency level to
pinpoint causes for the lack of competency. Further investigation determines the root causes of issues and establishes
safe operational zones for the module. If a non-safe region is identified, either the module is deactivated or must
undergo enhancement for continued use, requiring cautious consideration of decisions or results within the AI system.

\bigskip

In machine learning, constant evolution and non-deterministic changes in competency are inherent. AI systems with
multiple machine learning modules dynamically evolve their competency based on the context of the world model, sensed
data, and system accuracy levels. Consequently, the solution competency's evolution must align with individual modules'
evolving competencies. The critical challenge lies in explicitly defining solution competency for end-users to
comprehend and make informed decisions.

\bigskip

Formulating and computing micro-competency and macro-competency present inherent complexities. The inside-out paradigm
for AI system competency establishes the competency of modules and components based on their inherent capabilities,
determining the overall competency of the AI system. Conversely, the outside-in paradigm first establishes the
overarching competency of the AI system and then selects modules and components to meet these competency requirements.

\bigskip

Crucially, competency articulation at a detailed level and continuous reasoning during system operation are essential.
The enhancement of system competency relies on reflective processes conducted by the AI system. The system identifies
flaws in modules or components through ongoing audits, rectifying them to improve the overall competency. Explicit
articulation of micro-competency and macro-competency is the foundation for creating a robust framework for
contemplation, facilitating the continuous enhancement of the system's competency. Consequently, there is room for
meta-level processing to guide the system in contemplation and utilize feedback to improve its competency iteratively.

\section{Moral Competency}
In the present landscape, concerns loom over the potential for an AI system to exhibit rogue behavior, harming society.
Moral competency in this context implies a deficiency in the AI system's ability to align with its intended goals while
simultaneously possessing an acquired (or unintentional) competency to cause harm. The lack of competency often stems
from the inherently open nature of the operational world and gaps in competency at both micro and macro levels.
Consequently, the acknowledgment persists that the Glass Door Problem exists within AI systems, risking harm to
themselves or others. In specific scenarios, the AI system is compelled to act, and any chosen action results in harm,
with inaction also having detrimental consequences. Thus, a competent AI system should proactively avoid entering such
predicaments.

\bigskip

The nuanced examination of micro and macro competency, coupled with an audit trail documenting actions and their
contextual details, facilitates the identification of the sequence of events leading to morally challenging situations.
Consequently, a prospect exists for conducting competency-level analytics to discern cause-and-effect relationships.
Simultaneously, as the solution executes, preemptive measures are taken to prune actions that might lead to potential
moral dilemmas. The system leverages contemplation on prior solution runs to preprocess and enhance the competency of
modules and components. Therefore, a competency-driven articulation of the AI system offers both scope and direction in
tackling the issue of the system going rogue due to a deficiency in overall moral competency.

\section{Discussion}
The Glass Door problem, akin to a person (due to lack of competency) being distracted and inadvertently injuring
oneself, can manifest in AI systems when they become distracted or fragile due to environmental factors, context,
action outcomes, and sensed data. Deliberate distractions, such as introducing false sensor data, could exploit the
lack of competency in opposing AI systems, creating unintended consequences. With the proliferation of AI systems,
adversaries may exploit these vulnerabilities for strategic advantage. Meta-level contemplation-driven analytics and
audits can play a pivotal role in identifying and averting such possibilities.

\bigskip

The distinction between competency and system specifications (limitations and capabilities) is a crucial issue. System
specifications outline parameters within which the system can operate, while competency represents accumulated
knowledge about the interconnected capabilities of modules and components. For instance, a stemmer with 92\% accuracy
has competency defined by the circumstances under which it performs optimally. Competency evolves, necessitating a
meta-level guidance system to generate diverse scenarios for evaluating micro and macro-level competency.

\bigskip

Computational competency emerges as a promising research domain, compelling exploration of models, methods, and
procedures to gauge the competency of individual modules, components, and entire AI systems. Audit trails, competency
requirements specifications, and contemplation, facilitated by competency analytics, contribute to computational
competency. This process unveils insights into the moral and ethical functioning of the AI system.

\bigskip

An AI Systems requirement is the thorough evaluation and upgrading of the system under realistic and anticipated
scenarios to achieve the desired level of competency. Policymakers can mandate the execution of the system under
specific scenarios, akin to clinical trials conducted by regulatory bodies such as the FDA, accompanied by competency
analytics and a rigorous approval process for the production use of an AI system. Comprehensive documentation of these
trials should be subject to public scrutiny, paving the way for an open approval process for public AI system use. The
inherent value of a competent AI system can be leveraged to seek additional financial benefits for its implementers.

\section{Related Work}
(Horovitz 2016) at the US Senate hearing, the note stated the concept of 'Integrative Intelligence'
and 'democratization of AI' in the presentation of the AI system. Our AI system handles both these
aspects by considering an AI system as system-integrated software modules and hardware components, and open disclosure
of the system's competency is a move towards democratizing AI. The moral competency of robots, robots interacting with
humans, and multi-agent systems were proposed and studied by (Scheutz and Malle, forthcoming, Malle 2014; Mallem
Scheutz, Forlizzi, and Voiklis 2016; Scheutz 2016). The key aspects studied by them are to (i) present scenarios where
moral competency is apparent, (ii) bring in vocabulary and constructs to communicate about moral competency, (iii)
develop norms for formalizing moral competency, and (iv) empirically study nature of asymmetry in moral judgments.
(Brooks et al. 1998) present an architecture and key capabilities of an AI system having the essence of human
intelligence.

\bigskip

Our AI system, built using similar architecture, is useful in evaluating micro and macro-level competency. There is
still much work to be done to develop a framework that ensures the seamless integration of disparate and different AI
components from the perspective of competency analytics. (Scassellati 2002) delves into the nature of a robot when it
must interact with humans through sensors. Data and algorithmic competency issues occur because of sensor data
dependencies. (Domingos 2012) has clarified what a machine learning system is. Our use of clarity in representation
competency and other micro and macro competency aspects is based on it. (Asaro 2016) introduces the liability problem
but has yet to be able to formally reason about how to identify the cause of the liability systematically. Our notion
of micro and macro competency and relevant audit trails will provide a basis for judging the system's liability (in
terms of competency). (Kafah, Ajmeri, and Singh 2017) study normative multi-agent systems from the context of liveness
and safety that can contribute to the data competency of the AI system. (Rossi 2017, Conitzer et al. 2017) worked on
moral preferences and decision-making. However, articulating preference regarding values and pay-offs provides
competency in terms of the implicit and explicit rationale for selecting these values and then manipulating them to
decide on moral actions.

\bigskip

The system's competency is different from each instance of the working of the AI system, wherein different optimizations
could occur. Therefore, even though much work has been done at algorithmic or technique levels of moral and ethical
behavior, but still needs to develop the work to determine the accumulated competency from applying the methods and
techniques over many different scenarios.

(Gaur, Sangal, Bagaria 2010, chapter 8) presents how to articulate and reason about trust based on intention and
competency. Our approach for determining and reasoning about competency is along similar articulation of trust.

\bigskip

In (Johnson and Bullock, 2023), the authors present the notion of fragility in AI systems, especially those based on
neural networks. Fragility and competency are interrelated concepts, a fragile AI system implies an incompetent AI
system, as it can fail often. A competent AI system can become fragile and fail under certain circumstances. Fragility,
if it can be properly formulated and measured (like sensitivity studies in optimization problems (Yeung et al. 2010)),
can be used to quantify the level of competency of an AI system at a micro and macro level. 

\section{Conclusions}
AI systems, whether collaborating with humans, working autonomously, or supporting human activities, are increasingly
pervasive, eliciting apprehension due to concerns about potential harm. Humans, recognized for their competency, are
not immune to lapses, exemplified by the common occurrence of the {\textquotedbl}Glass Door problem.{\textquotedbl}
Analogously, AI systems inherently grapple with analogous challenges, necessitating a proactive approach to articulate
and address their competency.

\bigskip

The primary contribution of this note lies in advocating for a 'competency-oriented' paradigm during the design,
construction, and deployment of AI systems. This entails elevating competency to first-class status, whether at the
micro level within individual modules or components or the macro level, spanning the entire AI system. The critical
components that drive AI system functionality are examined to discern micro and macro-level competencies.

\bigskip

The objective is to enable competency-driven assessments by subjecting the AI system to diverse scenarios over extended
periods (contemplation). Employing competency analytics models supports subsequent computational evaluations, stating
and ensuring competency to prevent system-induced harm. Furthermore, these analytics serve as a valuable tool for
enhancing the system's competency.

\bigskip

Future work will focus on developing a comprehensive framework and a systematic approach to architect and implement a
competency-driven open AI system for deployment. Such a system aims to serve humanity benevolently, fostering
transparency, reliability, and ethically. 

\section*{References}
Scott V. Liebman, 404 S.W.2d 288(1966); Supreme Court of Texas, May 18, 1966
(https://law.justia.com/cases/texas/supreme-court/1966/a-10939-0.html).

Horvitz, E; 2016. Reflections on the Status and Future of Artificial Intelligence. \textit{Hearing on the Dawn of
Artificial Intelligence}.

Scheutz, M.; and Malle, B. F.; (forthcoming) \textit{Moral Robots.} in K. Rommelfanger and S. Johnson (eds.), Routledge
Handbook of Neuroethics. New York, NY: Routledge/Taylor \& Francis.

Malle, B.F.; (2014). \textit{Moral competence in robots?} In Seibt, J, Hakli, R., \& Norskov, M. (Eds), Social robots
and the future of social relations, Proceedings of Robo-Philosophy.

Malle, B. F.; Scheutz, M,; Forlizzi, J.; and Voiklis J. (2016) Which Robot am I Thinking About? The impact of Action and
Appearance on People's Evaluations of a Moral Robot. ACM/IEEE conference on Human and Robot
Interaction. 

Scheutz, M;. (2016) The need for moral competency in autonomous agent architectures. In Vincent C. Muller (Ed.)
Fundamentals of Artificial Intelligence. Springer.

Brooks, R. A.; Breazeal, C.; Irie, R.; Kemp, C. C.; Marjanovic, M.; Scassellati, B.; and Williamson, M.M.; 1998.
Alternative Essences of Intelligence. AAAI Conference Proceedings.

Scassellati, B.; (2002) Theory of Mind for a Humanoid Robot. Autonomous Robots.

Domingos, P.; (2012) A few useful things to know about machine learning. CACM pages 78-87.

Asaro, P. M; (2016) The liability Problem for Autonomous Artificial Agents. AAAI Symposium on Ethical and Moral
considerations in Non-Human Agents, Stanford University, Stanford, CA.

Kafah, O.; Ajmeri, N.; and Singh, M. (2017) KONT: Computing Tradeoffs in Normative Multiagent Systems. AAAI Conference.

Rossi, F.; (2016) Moral Preferences. 10\textsuperscript{th} Workshop on advances of Preference Handling.

Conitzer, V.; Sinnott-Armstrong, W.; Borg, J.S.; Deng, Y.; Kramer, M.; (2017) Moral Decision-Making Frameworks for
Artificial Intelligence. AAAI Conference. 

Gaur, R. R.; Sangal, R.; Bagaria, G.P.; (2010) A foundation course in Human Values and Professional Ethics. Excel Books.
(Available on Kindle).

Johnson, J. A., Bullock D. H.; (2023) Fragility in Ais using Artificial Neural Networks, CACM, July 2023.

Yeung D. S., Cloete I., Shi D., Ng W. W. Y.; (2010) Sensitivity Analysis for Neural Networks, Springer Verlag.





\end{document}